\documentclass[preprint,showpacs,amsmath,floatfix,prl,aps]{revtex4}
\usepackage{graphicx}

\begin{document}

\title{All-electron Exact Exchange Treatment of Semiconductors: Effect of 
Core-valence Interaction on Band-gap and $d$-band Position}

\author{S. Sharma}
\email{sangeeta.sharma@uni-graz.at}
\author{J. K. Dewhurst}
\author{C. Ambrosch-Draxl}
\affiliation{Institut f\"{u}r Physik, Karl--Franzens--Universit\"at Graz,
Universit\"atsplatz 5, A--8010 Graz, Austria}

\date{\today}

\begin{abstract}
Exact exchange (EXX) Kohn-Sham calculations within an all-electron full-potential
method are performed on a range of semiconductors and insulators (Ge, GaAs, CdS,
Si, ZnS, C, BN, Ne, Ar, Kr and Xe). We find that the band-gaps are not as close to
experiment as those obtained from previous pseudopotential EXX calculations. 
Full-potential band-gaps are also not significantly better for $sp$ semiconductors 
than for insulators, as had been found for pseudopotentials.
The locations of $d$-band states, determined using the full-potential EXX method, 
are in excellent agreement with experiment, irrespective of whether these states 
are core, semi-core or valence. We conclude that the inclusion of the core-valence 
interaction is necessary for accurate determination of EXX Kohn-Sham band 
structures, indicating a possible deficiency in pseudopotential calculations.
\end{abstract}

\pacs{71.15.Mb,71.10.-w,71.22.+i}

\maketitle

Exact treatment of exchange within the Kohn-Sham (KS) formulation of density
functional theory (DFT) has been one of the most interesting 
developments in the last few years 
\cite{gorling94,gorling96,stadele97,kotani95,magyar04}.
The KS band-gaps ($E_{\rm g}^{\rm KS}$) for $sp$ semiconductors, determined using
exact exchange (EXX) and the pseudopotential (PP) method, are found to be very 
close to experimental gaps \cite{stadele97,stadele99}, even in the absence of
any correlation. This is surprising because
the relation between $E_{\rm g}^{\rm KS}$ and the fundamental band-gap $E_{\rm g}$ of 
solids is given by $E_{\rm g}=E_{\rm g}^{\rm KS}+\Delta_{\rm xc}$ where,
$\Delta_{\rm xc}$ is the discontinuity in the exchange-correlation potential.
This discontinuity is lacking in the local density approximation (LDA),
resulting in the equality of the KS and fundamental band-gap \cite{magyar04} 
which leads to the famous band-gap problem.
But for the EXX potential even in the absence of exact correlation, 
the discontinuity does not vanish and hence one cannot expect that the EXX KS
gap alone can be identified directly with experimental values. In fact,
St{\"a}dele {\it et al.} \cite{stadele97} determined the discontinuities in the
exchange potential explicitly for Si, Ge and GaAs, and found them to be about
3-5 times the band-gap.
Furthermore, Magyar {\it et al.} \cite{magyar04} noted that the EXX KS gap for
wide band-gap noble
gas solids is not as close to the experimental gap (either optical or fundamental) as 
it is for $sp$ semiconductors. One of the major conclusions of this work was
that the EXX method does not provide a KS band structure that agrees equally well
with experiment for semiconductors and insulators.
All the above work has been done using the PP approach. This approach neglects
both the exact core-valence exchange interaction as well as the relaxation of the
core states in the solid-state environment.
There do exist all-electron EXX calculations of the band-gaps in C, Si and Ge
\cite{kotani95}. In this work, the linear combination of muffin-tin orbitals (LMTO)
within the atomic sphere approximation (ASA) used indicates
that the band-gaps are an improvement over LDA but do not agree with 
experiment as well as PP-EXX. These calculations however suffer from the shape
approximation of the
potential which ignores non-spherical contributions. In our recent article,
the importance of this non-spherical part of the self-consistent 
EXX potential was demostrated\cite{ss05}.

At this point it is relevant to note that another method used to improve upon 
the single particle KS band-gaps is the $GW$ approximation. Here a partially or
fully self-consistent solution of the Dyson equation is used to determine the 
self-energy which provides a correction to the KS-LDA band-gap. It has been inferred from 
$GW$ calculations performed using the PP approach that full self-consistency, even
though it is essential for charge conservation, worsens the agreement of 
quasi-particle band-gaps with experimental data \cite{holm98,schone98,holm99}. 
Very recently it was noted by Ku and Eguiluz \cite{ku02} that the success
of the partially self-consistent $GW$ approach is due to some kind of error cancellation
which is an artifact of the PP and also that the deep-core states have an important 
effect on the band-gaps. Such an effect may be of similar importance for EXX calculations
and requires thorough investigation.

Another important contributory factor to the gap error is the cation $d$-band 
position in $sp$ semiconductors, especially when these levels are relatively shallow 
(e.g. GaAs, ZnS, CdS) \cite{kotani02}. In these cases, use of the LDA leads to a 
misplacement of the $d$-bands, although this can be improved upon by the
$GW$ approximation or the self-interaction correction (SIC) to the potential
\cite{aryasetiawan96,vogel96,kotani02}. Previous PP-EXX results indicate that
even though the EXX potential is self-interaction free and the band-gaps are an 
improvement over those of LDA, the $d$-band positions are still not correct 
\cite{rinke}.

In this Letter, we present results of all-electron full-potential EXX calculations
applied to representative semiconductors and insulators. This is to ascertain
the effect of core-valence interaction and asymmetry in the potential on band-gaps
and $d$-band positions, and also to find the reason for the inconsistency in EXX
band-gaps of semiconductors and insulators.

We have implemented EXX potential using the all-electron full-potential
linearized augmented-plane wave (FP-LAPW) method \cite{singh} within the
{\sf EXCITING} code \cite{exciting}. The starting point for these calculations is the
exchange energy
$$ E_{\rm x}[n]=-\frac{1}{2}\sum_{i{\bf k},j{\bf k}'}^{\rm occ}w_{\bf k}w_{{\bf k}'}
 \int d{\bf r}\;d{\bf r}'
 \frac{\phi_{i{\bf k}}^*({\bf r})\phi_{j{\bf k}'}^*({\bf r}')\phi_{j{\bf k}'}({\bf r})
 \phi_{i{\bf k}}({\bf r}')}{|{\bf r}-{\bf r}'|}, $$
where $\phi_{i{\bf k}}$ is the $i$th Kohn-Sham orbital of $k$-point ${\bf k}$,
$w_{\bf k}$ is the $k$-point weight,
$n({\bf r})\equiv\sum_{i{\bf k}}^{\rm occ}w_{\bf k}\phi_{i{\bf k}}^*({\bf r})
\phi_{i{\bf k}}({\bf r})$ is the density, and $i$ and $j$ run over
both core and valence states, in contrast to PPs where the indices run only over
the valence states. Proceeding in the same manner as G\"{o}rling {\it et al.}
\cite{gorling94,gorling96}, the functional derivative chain rule is employed to
obtain the exchange potential
\begin{align}\label{eq1}
 v_{\rm x}[n]({\bf r})&\equiv\frac{\delta E_{\rm x}[n]}{\delta n({\bf r})}\nonumber\\
 &=\sum_{i{\bf k}}^{\rm occ}w_{\bf k}\int d{\bf r}'\;d{\bf r}''\left[
 \frac{\delta E_{\rm x}}{\delta \phi_{i{\bf k}}({\bf r}'')}
 \frac{\delta \phi_{i{\bf k}}({\bf r}'')}{\delta v_{\rm s}({\bf r}')}+
 \frac{\delta E_{\rm x}}{\delta \phi_{i{\bf k}}^*({\bf r}'')}
 \frac{\delta \phi_{i{\bf k}}^*({\bf r}'')}{\delta v_{\rm s}({\bf r}')}\right]
 \frac{\delta v_{\rm s}({\bf r}')}{\delta n({\bf r})}\nonumber\\
 &=\int d{\bf r}'\left[\sum_{i{\bf k}}^{\rm occ}\sum_{j}^{\rm unocc}
 w_{\bf k}\langle\phi_{i{\bf k}}|\hat{v}^{\rm NL}_{\rm x}|\phi_{j{\bf k}}\rangle
 \frac{\phi^*_{j{\bf k}}({\bf r}')\phi_{i{\bf k}}({\bf r}')}
 {\varepsilon_{i{\bf k}}-\varepsilon_{j{\bf k}}}+{\rm c.c.} \right]
 \frac{\delta v_{\rm s}({\bf r}')}{\delta n({\bf r})},
\end{align}
where $\varepsilon_{i{\bf k}}$ are the Kohn-Sham eigenvalues and 
\begin{align}\label{eq2}
\langle\phi_{i{\bf k}}|\hat{v}^{\rm NL}_{\rm x}|\phi_{j{\bf k}}\rangle=
 \sum_{l{\bf k}'}^{\rm occ}w_{{\bf k}'}\int d{\bf r}\;d{\bf r}'
 \frac{\phi_{i{\bf k}}^*({\bf r})\phi_{l{\bf k}'}^*({\bf r}')
 \phi_{l{\bf k}'}({\bf r})\phi_{j{\bf k}}({\bf r}')}{|{\bf r}-{\bf r}'|}.
\end{align}
In order to obtain the functional derivative
$\delta v_{\rm s}({\bf r}')/\delta n({\bf r})$, the linear-response operator $\chi$,
defined by the relation
$$ \delta n({\bf r})=\int d{\bf r}'\;\chi({\bf r},{\bf r}')\delta v_{\rm s}({\bf r}'), $$
must be inverted. This is not possible directly as $\chi$ has a zero eigenvalue
corresponding to a constant eigenfunction. However, in a basis which excludes such
functions the inversion of $\chi$ can be performed. Application of elementary perturbation
theory to $n$ leads to an explicit expression for the response
$$ \chi({\bf r},{\bf r}')=\sum_{i{\bf k}}^{\rm occ}\sum_{j}^{\rm unocc}
 w_{\bf k}\frac{\phi_{i{\bf k}}^*({\bf r})\phi_{j{\bf k}}({\bf r})
 \phi_{j{\bf k}}^*({\bf r}')\phi_{i{\bf k}}({\bf r}')}
 {\varepsilon_{i{\bf k}}-\varepsilon_{j{\bf k}}}+{\rm c.c.} $$
Implementations of EXX for pseudopotentials have used the plane wave basis for
expanding $\chi$. This is not feasible in a FP-LAPW approach because of the
strongly varying wave functions inside the muffin-tin. Instead, we propose a new
basis as follows. The overlap
densities $\rho_{\alpha}({\bf r})\equiv\phi_{i{\bf k}}^*({\bf r})\phi_{j{\bf k}}({\bf r})$
and their complex conjugates are used as a spanning set, where
$\alpha\equiv(i{\bf k},j{\bf k})$ with $i$
and $j$ labelling occupied and unoccupied states, respectively. These are the states of
some representative $k$-point which is chosen in advance.
This spanning set is then reduced to a basis set as follows. First the
overlap matrix of the spanning set elements
$$ O_{\alpha\beta}\equiv\int d{\bf r}\;\rho_{\alpha}^*({\bf r})\rho_{\beta}({\bf r}) $$
is obtained and diagonalized. Note that $O$, which is positive semidefinite,
has nonnegative eigenvalues.
Next, all the eigenvectors of $O$ corresponding to eigenvalues smaller than
a certain tolerance, $\epsilon$, are discarded. Finally, denoting the remaining
eigenvectors as $v^{\beta}_{\gamma}$, where $\beta$ labels the vector and $\gamma$
its coordinate, we seek a transformation matrix $C$ such that if
$$ \tilde{\rho}_{\alpha}({\bf r})=\sum_{\beta}C^{\alpha}_{\beta}\sum_{\gamma}
 v^{\beta}_{\gamma}\rho_{\gamma}({\bf r}) $$
then
$$ \int d{\bf r}\;\tilde{\rho}_{\alpha}^*({\bf r})\tilde{\rho}_{\beta}({\bf r})
 =\delta_{\alpha\beta}. $$
Straight-forward algebra shows that $C$ should satisfy
\begin{equation}\label{eq3}
 CC^{\dag}=\left(v^{\dag}Ov\right)^{-1}
\end{equation}
where $v$ is the matrix of eigenvectors $v^{\beta}_{\gamma}$ in column-wise form.
The matrix $C$ is obtained by performing a Cholesky decomposition on the right hand
side of Eq. \ref{eq3}. By virtue of their construction and the fact that
$\int d{\bf r}\;\rho_{\alpha}({\bf r})=0$, the set of functions
$\{\tilde{\rho}_\alpha\}$ form an optimal basis for the expansion and inversion of
$\chi$, as well as for the term in square brackets in Eq. \ref{eq1}. We should also
point out that this basis may be useful for FP-LAPW time-dependent DFT response
and $GW$ methods.

Special attention should now be drawn to the calculation of the non-local matrix 
elements (Eq. \ref{eq2}) which may be determined by a well-established method for 
solving Poisson's equation in a FP-LAPW environment \cite{singh}. The differences 
here are that the densities are now complex and there is a long-range term arising 
when ${\bf q}\equiv{\bf k}-{\bf k}'$ is close to zero. This is treated by 
considering the so-called pseudo-charge density \cite{singh}, which is chosen to 
be sufficiently smooth within the muffin-tins so that it may be expanded in terms 
of plane waves, and yet has the same multipole expansion as the real density. 
If we restrict ${\bf k}$ and ${\bf k}'$ such that ${\bf q}$ is in the first 
Brillouin zone (BZ), then the long-range (LR) contribution to the matrix elements 
from the pseudo-charge is
$$ \langle\phi_{i{\bf k}}|\hat{v}^{\rm NL}_{\rm x}|\phi_{j{\bf k}}\rangle_{\rm LR}=
 \sum_{l{\bf q}}^{\rm occ}w_{\bf q}\frac{4\pi\Omega}{q^2}\rho_{il}^*({\bf q})
 \rho_{lj}({\bf q}), $$
where $\rho_{il}({\bf q})$ and $\rho_{lj}({\bf q})$ are the pseudo-charge densities
in reciprocal space.
This sum suffers from poor convergence with respect to the number of $q$-points.
We therefore approximate it by an integral over a sphere of volume equivalent to
that of the BZ. The final expression is
$$ \langle\phi_{i{\bf k}}|\hat{v}^{\rm NL}_{\rm x}|\phi_{j{\bf k}}\rangle_{\rm LR}
 \simeq 2\left(\frac{6\Omega^5}{\pi}\right)^{1/3}
 \sum_{l{\bf q}}^{\rm occ}w_{\bf q}\rho_{il}^*({\bf q})\rho_{lj}({\bf q}). $$
All the calculations in the present work are performed on a mesh of 27 special
$k$-points in the irreducible BZ. As the response and its inverse
are very sensitive to the
number of empty states, we have taken special care to use a sufficient number to
converge the band-gaps to within 0.01 eV. We find that almost all the materials
studied here have converged $d$-band positions and band-gaps with around 25 empty
states. All the results presented are generated using 30 empty states. One final
practical note is that the eigenvalue cut-off, $\epsilon$, discussed above, is 
taken to be $10^{-5}$ times the largest eigenvalue of $O$. For example for Si this 
results in 73 elements in the full basis set.

\begin{figure}[ht]
\centerline{\includegraphics[width=0.8\columnwidth,angle=-90]{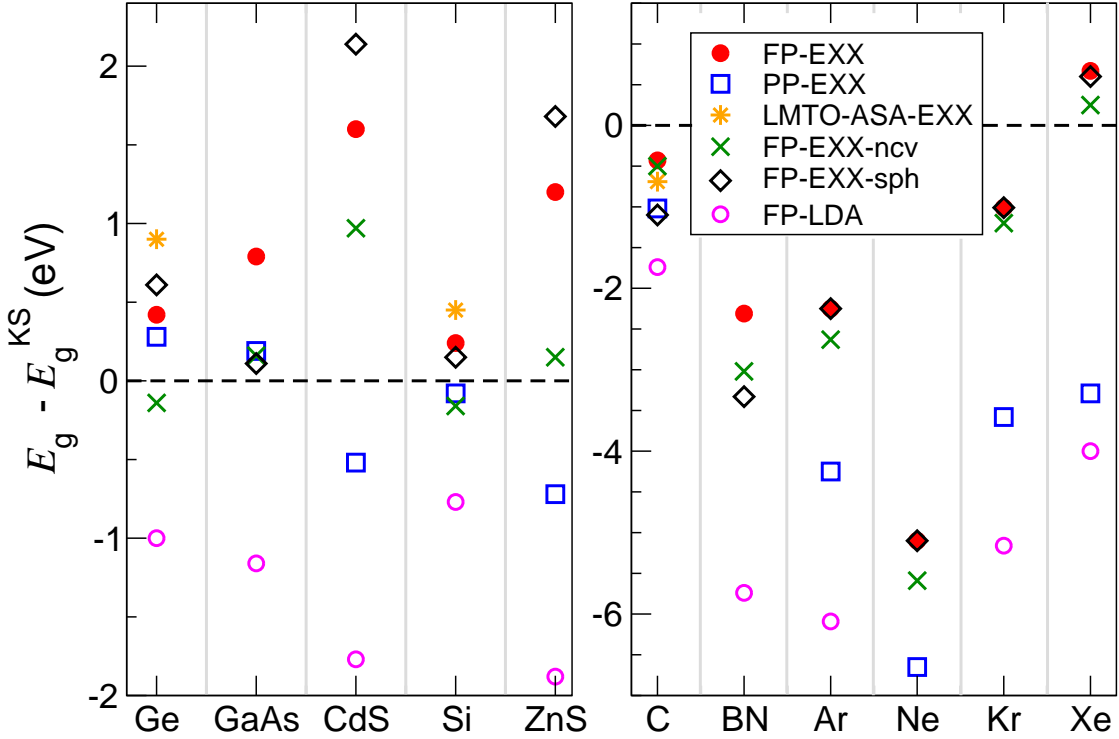}}
\caption{Difference between theoretical KS gaps and experimental fundamental gaps in eV.
Full-potential exact exchange (FP-EXX), FP-EXX without core-valence interaction
(FP-EXX-ncv), FP-EXX with only spherical exchange (FP-EXX-sph) and
LDA (FP-LDA) results are from this work. The PP-EXX results are from Refs.
\onlinecite{stadele99,magyar04,rinke}. The LMTO-ASA-EXX results are from Ref.
\onlinecite{kotani95} and the experimental data is taken from Refs.
\onlinecite{expt1,expt2,zahn94}. Note that LDA gap difference for Ne
is -10.25 eV, and is off the scale of the plot.}\label{fig1}
\end{figure}

In Fig. 1 we present the difference between the experimental fundamental band-gap and 
the KS band-gap for various semiconductors and insulators. 
The EXX potential is self-interaction free and hence an enhancement of the 
KS band-gap over that of the LDA is expected. This is indeed the case here.
We find that for semiconductors the FP-EXX band-gaps are overestimated with 
respect to experiments by up to 60\%.
The band-gaps in all the insulators except Xe are underestimated with respect to 
the experimental fundamental gap up to about 24\%.  For Kr and Xe the KS band-gap 
is overestimated with respect to the experimental optical gap by 3.8\% and 24.6\%
respectively, while for other insulators it is underestimated up to 6\% .

Unlike for PP-EXX, there is not the excellent agreement of the KS gaps
of $sp$ semiconductors with experiment. In contradistinction to the PP-EXX 
results we find that the EXX is not inconsistent in its treatment of semiconductors
and insulators, in fact agreement with experiment of the KS gaps for insulators
is not any worse than for the semiconductors.
In order to determine the reason for this behaviour we performed ground state 
calculations for these materials by ignoring the core-valence interaction term in 
the EXX potential. We emphasise here that this term was removed only from the
exchange potential and not from the rest of the potential. Even this has substantial
effect on the KS band-gaps. The results are marked as FP-EXX-ncv in Fig. 1. Compared
to FP-EXX, the FP-EXX-ncv band-gaps for all the semiconductors are closer to the 
PP-EXX and the experimental values. 
On the other hand, the agreement of the KS band-gaps with experiments for insulators now
worsens. This trend is concomitant with the findings of the PP-EXX calculations
\cite{stadele99,magyar04} where the performance of the EXX potential in determining 
the KS gaps is excellent for the semiconductors and lacking for the insulators.
Previous all electron EXX calculations performed within the ASA
show a similar overestimation of the gaps for semiconductors \cite{kotani95}. 
All this suggests that the lack of core-valence interaction is responsible 
for the spurious agreement of the PP-EXX KS band-gaps with the experiments for $sp$
semiconductors.

In our work on magnetic metals \cite{ss05} we found that inclusion of the 
non-spherical contributions to the exchange potential is crucial to obtain the 
correct ground-state. The results for the KS band-gaps obtained 
using  only the spherical contribution to the EXX potential are marked as 
FP-EXX-sph in Fig. 1. The effect of the shape approximation to the potential 
in all the materials other than noble gas solids is substantial. Which is 
expected since the noble gas solids are composed of almost independent atoms.

\begin{figure}[ht]
\centerline{\includegraphics[width=0.8\columnwidth,angle=-90]{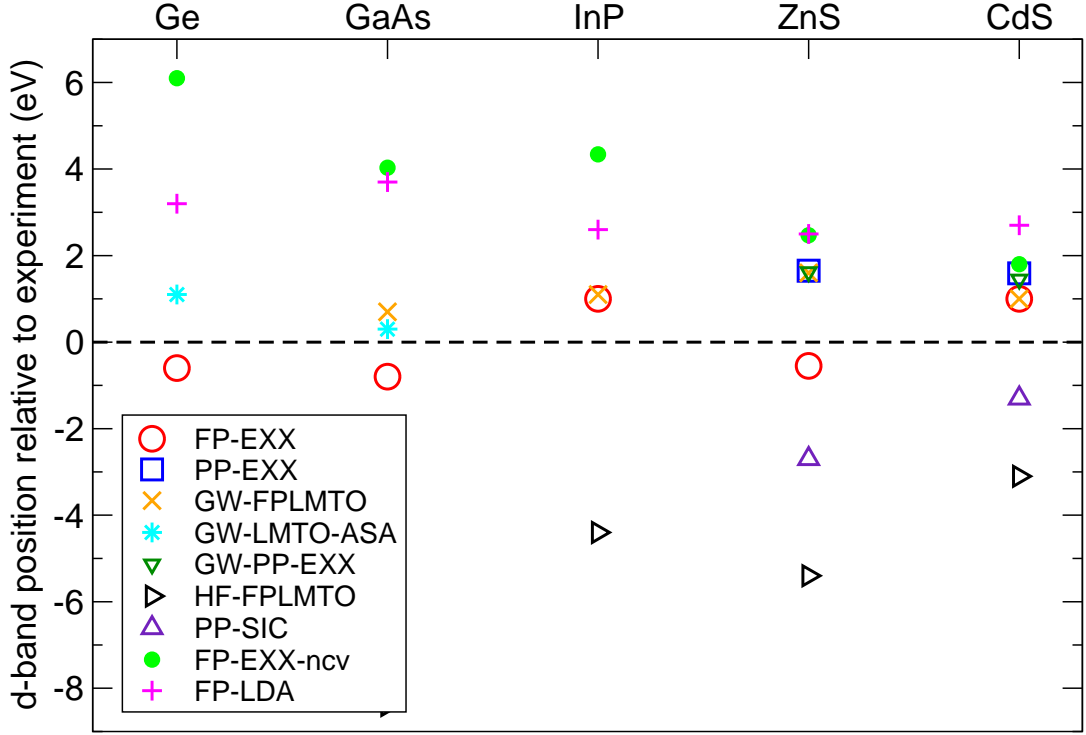}}
\caption{Difference between theoretical $d$-band eigenvalues and experimental
data in eV. Full-potential exact exchange (FP-EXX), FP-EXX without core-valence
interaction (FP-EXX-ncv) and LDA (FP-LDA) results are from this work.
The results obtained using the $GW$ calculation on top of the PP-EXX ground-state
($GW$-PP-EXX) are from Ref. \onlinecite{rinke}, the self-interaction corrected 
potential within PP approach (PP-EXX) are from Ref. \onlinecite{vogel96}. 
$GW$ results calculated using a FP-LMTO method ($GW$-FPLMTO) and Hartree-Fock 
FP-LMTO (HF-FPLMTO) are from Ref. \onlinecite{kotani02}. The $GW$ calculation on 
top of the LMTO-ASA-LDA ground-state ($GW$-LMTO-ASA) are from Ref. 
\onlinecite{aryasetiawan96}.}\label{fig2}
\end{figure}

Another property which is known to lead to errors in the band-gaps \cite{kotani02} 
and for which the LDA performs badly is the position of the semi-core $d$-states. 
The LDA under-binds leading to displacement of $d$-bands much above the experimental 
value.
In the past, a quasi-particle bandstructure was determined using the self energy 
calculated with the $GW$ approximation to correct the $d$-band position 
\cite{aryasetiawan96}. Also the SIC to the potential was found to lead to good 
agreement between the theoretical and experimental position of these $d$-bands. 
Since the EXX is self-interaction free in nature, it should improve
the $d$-band positions. Counter intutively Rinke {\it et al.} \cite{rinke} 
found that even though the PP-EXX calculations result in a downward motion of the 
$d$-bands with respect to LDA, still $GW$ calculations are needed to fully correct 
the positions. 
The FP-EXX potential, in which the core and valence states are treated on 
the same footing, is truly self-interaction free in nature and it would be 
interesting to see how this effects these semi-core/core-like $d$-bands. 

In the present work, we compare FP-EXX $d$-band eigenvalues with those of previous 
PP-EXX, experiment and various $GW$ and SIC calculations in Fig. 2. The compounds 
are chosen so that $d$-bands are in various energy regimes to get an overall
picture: core-like (Ge: experimentally -29.6 eV from the top of the valence band), 
semi-core (GaAs: -18.8 eV, InP: -16.8 eV) and valence (ZnS: -8.7 eV, CdS: -9.2 eV).
As can be seen, the eigenvalues of $d$-states obtained using the FP-EXX method are 
in excellent agreement with experiments. 
In fact, in several cases it is better than that of the FP-$GW$ results. 

Here again the agreement between the FP-EXX-ncv and PP-EXX values of Rinke 
{\it et al.} \cite{rinke} is quite good. This indicates that the inclusion of 
core-valence interaction is crucial for determination of the correct $d$-band 
eigenvalues.

The main aim of our paper is to establish the effects of the inclusion of core-valence
interaction in full-potential EXX calculations.
Prior to this investigation, it appeared that EXX might be the answer to the
band-gap problem at least for $sp$ semiconductors.
We find that this is not the case if core-valence interaction is included in the
calculation of the EXX potential. We also find that EXX is consistent in its 
treatment of semiconductors and insulators.
Our findings indicate a possible shortcoming in the PP-EXX method arising
from the lack of explicit core-valence interactions in the PP-EXX potential, as 
mentioned earlier by Ku and Eguiluz \cite{ku02}.
Furthermore, we show that the $d$-band eigenvalues determined using the
FP-EXX method are in 
excellent agreement with experiment irrespective of whether these states are in the
core, semi-core or valence. In most cases the agreement is better than that
achieved by the full-potential $GW$ calculations.

We acknowledge the Austrian Science Fund (project P16227) and the EXCITING 
network funded by the EU (Contract HPRN-CT-2002-00317) for financial support.

\end{document}